\documentclass[12pt]{article}
\usepackage{amssymb,latexsym,epsfig,cite}

\newcommand{\be}{\begin{equation}}
\newcommand{\ee}{\end{equation}}

\def\bea{\begin{eqnarray}}
\def\eea{\end{eqnarray}}
\begin{document}

\thispagestyle{empty}

\begin{center}
{\Large Beyond conventional factorization:
Non-Hermitian\\[1.5ex] 
Hamiltonians with radial oscillator spectrum}
\end{center}

\vskip2ex

\begin{center}
Ivan Cabrera-Munguia and Oscar Rosas-Ortiz\\[2ex]
{\footnotesize Departamento de F\'{\i}sica, Cinvestav, AP 14-740, 
07000 M\'exico DF, Mexico}
\end{center}

\begin{abstract}
\noindent
The eigenvalue problem of the spherically symmetric oscillator
Hamiltonian is revisited in the context of canonical raising and
lowering operators. The Hamiltonian is then factorized in terms of
two not mutually adjoint factorizing operators which, in turn,
give rise to a non-Hermitian radial Hamiltonian. The set of
eigenvalues of this new Hamiltonian is exactly the same as the
energy spectrum of the radial oscillator and the new
square-integrable eigenfunctions are complex Darboux-deformations
of the associated Laguerre polynomials.
\end{abstract}

%%----------------------------------------------------->{begin body}

\section{Introduction}

Factorization is a powerful method to compute eigenvalues and
recurrence relations for solutions of differential equations in
mathematical physics (see e.g.
\cite{Mil68,del91,Neg00a,Neg00b,Dong07}). Introduced in Quantum
Mechanics (QM) since Fock, Schr\"odinger and Dirac times, some of
the antecedents of this algebraic method can be found in the
geometrical formalism of Darboux and B\"acklund. It is also
notable that the (canonical) Fock-Dirac ladder operators have been
of big influence in the development of contemporary Physics over
the years \cite{Mie04}. The method, as formalized by Infeld and
Hull \cite{Inf51}, was extended by Mielnik to embrace factorizing
operators of the harmonic oscillator Hamiltoninan which do not
create nor annihilate occupation numbers in the Fock states
\cite{Mie84}. Applications of the Mielnik's method to radial
problems were reported first by Fern\'andez \cite{Fer84} and
connections with the Supersymmetric formalism of QM were found by
Nieto \cite{Nie84} and Sukumar \cite{Suk85a}. The relationship
between factorizing and Darboux-deforming potentials in QM was
extensively studied by Andrianov and coworkers \cite{And84a}.
Nowadays, the factorization method is of the highest utility in
constructing new exactly solvable potentials in QM (see e.g.
\cite{del02,susy01} and references quoted therein). In particular,
recent results show that the method can be used in solving the
problem of complex potentials with real spectrum
\cite{Bag95,Bay96,And99,Bag01,Fer03,Ros03} as well as in the study
of resonances and Gamow-Siegert states \cite{Ros07,Fer07a,Fer07b}.

In the one-dimensional case there is a unique expression for the
canonical ladder operators since the energy levels are not
degenerated. For instance, the Fock-Dirac operators transform an
eigenfunction of the harmonic oscillator into another one and
factorize, at the same time, the corresponding Hamiltonian.
However, in a general situation, the pair of operators which
factorize a given Hamiltonian do not necessarily correspond to the
shift operators. Distorted versions of the harmonic oscillator,
for example, require higher order raising and lowering operators
although the Hamiltonian is factorized by first order differential
operators \cite{Mie84,Fer94}. On the other hand, it is usual to
find degenerate energy levels in higher dimensions and several
kinds of shift operators are necessary to intertwine the
wave-functions \cite{del91}. A particular case is of special
interest: spherical symmetry in central potentials induces energy
degeneracies under rotations (different orientations of the
angular momentum vector lead to the same energy). Hence, the
ladder operators are labelled by the azimuthal quantum number
$\ell$ and their action mainly affects the $\ell$--dependence of
the solutions (see for instance
\cite{Inf51,Fer84,New78,Don87,Ros98}). Thus, they intertwine
solutions of potentials which differ in one unit of the angular
momentum.

In this work we are interested in the three dimensional isotropic
oscillator. The Hamiltonian of this system can be factorized in
four different forms by means of a basic set of canonical ladder
operators \cite{Fer96}. Each of these products preserves the
adjointness of the Hamiltonian since the factors are mutually
adjoint. In contrast, we are going to express the radial
oscillator Hamiltonian as the product of two not mutually adjoint
factorizing operators. The advantage of these new operators is
that they give rise to non-Hermitian radial Hamiltonians for which
the point spectrum is exactly the same as the spherically
symmetric oscillator one.

In Section~2 conventional factorizations of the isotropic
oscillator are revisited. Explicitly derived, the four canonical
ladder operators are used to construct the basis of physical
wave-functions (orthonormal, associated Laguerre polynomials) and
the energy spectrum. For a given $\ell$, the creation or
annihilation of nodes in the wave-functions require second order
differential operators which do not factorize the Hamiltonian. In
Section~3 a pair of not mutually adjoint factorizing operators is
introduced to extend the number of factorizations in which the
radial oscillator Hamiltonian can be expressed. These same
operators are used to construct a new Hamiltonian which is not
self adjoint under the inner product of the Laguerre--Hilbert
Space ${\cal H}$. The relevant aspect of our results is that the
eigenfunctions of the new Hamiltonian are square-integrable in
${\cal H}$. It is also introduced a pair of ladder operators which
increase or decrease the energy of this new system. As we shall
see, these shift operators are necessarily fourth order
differential operators. Finally, the paper is closed with some
concluding remarks in Section~4.

%%----------------------------------------------------->{Section}
\section{Radial Oscillator Revisited}

\vskip12pt

The time-independent Schr\"odinger equation for the isotropic
oscillator $V(r)=r^2$, in appropriate units and after separation
of variables, reduces to
\begin{equation}
H_{\ell} \, \phi(r,\ell) =E(\ell) \, \phi(r,\ell)
\label{schro1}
\end{equation}
where the azimuthal quantum number $\ell$ is a non-negative
integer, $E(\ell)$ is twice the dimensionless energy eigenvalue
and the radial Hamiltonian reads:
\begin{equation}
H_{\ell} \equiv - \frac{d^2}{dr^2} + V_{\ell}(r) =
-\frac{d^2}{dr^2} + \frac{\ell(\ell +1)}{r^2} + r^2.
\label{ham1}
\end{equation}
The effective potential $V_{\ell}(r)$ has the domain $D_V=[0,
\infty)$ and the functions $\phi (r, \ell)$ are defined in terms
of the usual radial wave-function $\phi (r, \ell) = r R(r, \ell)$.
To be physically interpretable, these last functions have to
satisfy $\int^{+\infty}_0 \vert R(r, \ell) \vert^2 r^2 dr <
\infty$. In the sequel, whenever there is no confusion we shall
use the shortcut notation $f \equiv
f(r,\ell,\varepsilon_{\ell},\dots)$, keeping implicit the
dependence of $f$ on $r$, $\ell$, $\varepsilon_{\ell}$, and other
possible variables and parameters.

Conventional factorization is useful in solving equation
(\ref{schro1}) by expressing (\ref{ham1}) as the product of two
mutually adjoint first order differential operators
\be
H_{\ell} = a^{\dagger}_{\ell} \, a_{\ell} + \varepsilon_{\ell};
\qquad a^{\dagger}_{\ell}:= -\frac{d}{dr} + \alpha(r,\ell); \qquad
(a^{\dagger}_{\ell})^{\dagger} = a_{\ell}
\label{factor1}
\ee
with $\varepsilon_{\ell}$ a real ({\it factorization\/}) constant
to be fixed and $\alpha(r,\ell)$ a function satisfying the Riccati
equation
\be
-\alpha' + \alpha^2 = V_{\ell} - \varepsilon_{\ell}.
\label{rica1}
\ee
A particular solution of equation (\ref{rica1}) is immediate:
\be
\alpha = r -\frac{\ell +1}{r}, \qquad \varepsilon_{\ell} = 2 \ell
+ 3.
\label{rica1a}
\ee
This {\it superpotential\/} $\alpha$ leads to one of the canonical
forms of the {\it factorizing\/} operators:
\be
a_{\ell} = \frac{d}{dr} -\frac{\ell + 1}{r} + r, \qquad
a^{\dagger}_{\ell} = -\frac{d}{dr} -\frac{\ell + 1}{r} + r.
\label{factor1a}
\ee
The first advantage of the method is clear by noticing that a
solution of $a_{\ell} \, \phi_0(r, \ell)= 0$ gives a solution of
(\ref{schro1}) belonging to the eigenvalue $E_0(\ell):=
E^{(\ell)}_0 =\varepsilon_{\ell}$. The calculation leads to
\be
\phi_0(r,\ell):= \phi^{(\ell)}_0(r) = C^{(\ell)}_0 \, r^{\ell +1}
e^{-r^2/2}
\label{ground}
\ee
which is free of nodes and becomes zero at the edges of $D_V$.
Hence, $C^{(\ell)}_0$ can be taken as the normalization constant
and, by applying the Sturm oscillation theorem \cite{Ber91}, we
know that $E^{(\ell)}_0 = 2\ell +3$ is the ground state energy of
$H_{\ell}$. Now, it is useful to reverse the product
(\ref{factor1}) to get
\be
H_{\ell +1} = a_{\ell} \, a^{\dagger}_{\ell} +
\varepsilon_{\ell}-2:= a_{\ell} \, a^{\dagger}_{\ell} +
\theta_{\ell}.
\label{factor2a}
\ee
Notice that the azimuthal quantum number $\ell$ plays the role of
a parameter in all the previous expressions. Hence, by changing
$\ell \rightarrow \ell-1$ and $\ell \rightarrow \ell +1$ in
(\ref{factor2a}) and (\ref{factor1}) respectively we also obtain
\be
H_{\ell} = a_{\ell -1} a^{\dagger}_{\ell -1} + \theta_{\ell -1},
\qquad H_{\ell +1} =  a^{\dagger}_{\ell +1} \, a_{\ell +1} +
\varepsilon_{\ell +1}.
\label{factor2b}
\ee
As a consequence, the following intertwining relationships are
true
\be
a_{\ell} \, (H_{\ell} -2) = H_{\ell+1} \, a_{\ell}, \qquad
a^{\dagger}_{\ell-1} \, (H_{\ell} +2) = H_{\ell-1} \,
a^{\dagger}_{\ell-1}.
\label{intertwin1}
\ee
Thereby, if $\phi(r,\ell)$ is eigenfunction of $H_{\ell}$ with
eigenvalue $E(\ell)$, then $\phi(r,\ell +1) \propto a_{\ell}
\,\phi (r,\ell)$ and $\phi(r,\ell -1) \propto a^{\dagger}_{\ell
-1} \,\phi (r,\ell)$ respectively satisfy the eigenvalue equation
of $H_{\ell +1}$ and $H_{\ell -1}$ with eigenvalue $E(\ell +1) =
E(\ell) -2$ and $E(\ell -1) = E(\ell) +2$.

Let us take full advantage of these last results. After applying
the R.H.S. relationship (\ref{intertwin1}) on the ground state
$\phi^{(\ell)}_0$, we arrive at the function
\be
\phi^{(\ell-1)}_1 \propto a^{\dagger}_{\ell -1} \, \phi^{(\ell)}_0
= (-2r^2 + 2\ell +1) \, \phi^{(\ell-1)}_0
\label{sol1}
\ee
which has a single node at $r=\sqrt{(2\ell +1)/2} \in D_V$ and
satisfies the Schr\"odinger equation
\be
H_{\ell -1} \, \phi^{(\ell-1)}_1 = (E^{(\ell)}_0 + 2)\,
\phi^{(\ell-1)}_1.
\label{schro3}
\ee
Function (\ref{sol1}) can be rewritten as
\be
\phi^{(\ell-1)}_1 \propto \phi^{(\ell-1)}_0 \, \left\{ r^{-2(\ell
-1/2)} \, e^{r^2} \left( \frac{1}{2r} \frac{d}{dr} \right)
r^{2(\ell + 1/2)} \, e^{-r^2} \right\}.
\label{sol1a}
\ee
After an intermediary change of independent variable $r^2 =x$, one
arrives at
\be
\phi^{(\ell-1)}_1 (r)= C^{(\ell-1)}_1 \, \phi^{(\ell-1)}_0 (r)\,
L^{(\ell -1) + 1/2}_1 (r^2)
\label{sol1b}
\ee
where $C^{(\ell-1)}_1$ is a normalization constant and
\be
L^{\nu}_n(x) = \frac{1}{n!} x^{-\nu} \, e^x \, \frac{d^n}{dx^n}
(e^{-x} x^{n+ \nu}), \qquad \nu >-1
\label{laguerre}
\ee
is the associated Laguerre polynomial of degree $n$. A similar
procedure leads to
\be
\phi^{(\ell-2)}_2 (r)= C^{(\ell-2)}_2 \, \phi^{(\ell-2)}_0 (r) \,
L^{(\ell -2) + 1/2}_2 (r^2) \propto a^{\dagger}_{\ell -2} \,
\phi^{(\ell-1)}_1 (r)
\label{sol1c}
\ee
which is the eigenfunction of $H_{\ell -2}$ with eigenvalue
$E^{(\ell-2)}_2 = E^{(\ell-1)}_1 +2 = E^{(\ell)}_0 + 4$. The
procedure can be applied $s$ times to get the eigenfunction of
$H_{\ell -s}$ with eigenvalue $E^{(\ell-s)}_s = E^{(\ell)}_0 +
2s$:
\be
\phi^{(\ell-s)}_s =  C^{(\ell-s)}_s \left(\prod_{k=0}^{s-1}
a^{\dagger}_{\ell-s+k} \right) \phi^{(\ell)}_0 := C^{(\ell-s)}_s
\left( a^{\dagger}_{\ell-s} \cdots a^{\dagger}_{\ell -2} \,
a^{\dagger}_{\ell-1} \right) \phi^{(\ell)}_0, \quad s = 1,2,\ldots
\label{sol1d}
\ee
Changing $\ell \rightarrow \ell+s$ and using $L^{\nu}_0 \equiv 1$,
this last expression leads to
\be
\phi^{(\ell)}_s (r)  = C^{(\ell)}_s \, \phi^{(\ell)}_0(r) \,
L^{\ell +1/2}_s (r^2), \qquad s = 0,1,\ldots
\label{sol1e}
\ee
Function (\ref{sol1e}) is the physical solution of (\ref{schro1})
with eigenvalue $E^{(\ell)}_s = E^{(\ell + s)}_0 +2s
=\varepsilon_{\ell + s} +2s = 2(2s + \ell) +3$. The presence of
nodes in $\phi^{(\ell)}_s$ is due to the zeros of the polynomial
$L^{\ell +1/2}_s$, the number of which is equal to $s$. By taking
$n = 2s + \ell = 1,0, \ldots$, we can use the conventional
notation $\phi^{(\ell)}_s (r) \equiv \phi_n(r)$ and $E^{(\ell)}_s
\equiv E_n = 2n+3$.

\medskip
%%%%%%%%%%%%%%%%%%%%%%%%%%%%%%
\begin{figure}[ht]
\begin{center}
\includegraphics[height=.25\textheight]{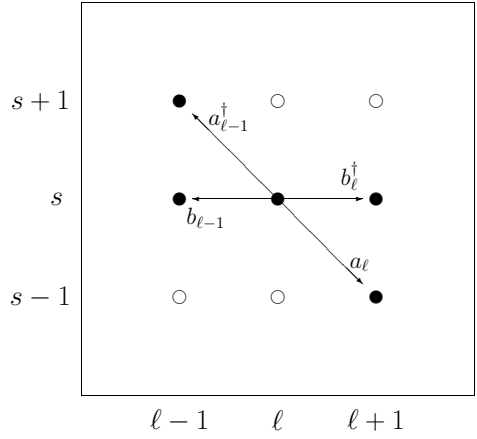}
\caption{\label{plano} Lattice representation in the
$(s\times\ell)$--plane for the action of the four canonical
factorization operators $a_{\ell}$, $a^{\dagger}_{\ell}$,
$b_{\ell}$, $b^{\dagger}_{\ell}$, of the radial oscillator.}
\end{center}
\end{figure}
%%%%%%%%%%%%%%%%%%%%%%%%%%%%%%

\noindent
Remark that the action of $a_{\ell}$ on an arbitrary function
(\ref{sol1e}), different from $\phi^{(\ell)}_0$, produces
$a_{\ell} \, \phi^{(\ell)}_s = \phi^{(\ell+1)}_{s-1}$. Thus
$a_{\ell}$ annihilates one node of the wave-function at the cost
of increasing in one unit the azimuthal quantum number $\ell$.
This can be represented in the Cartesian $(s\times\ell)$--plane as
the diagonal displacement $(s,\ell) \rightarrow (s-1,\ell+1)$; a
schematic picture of which is shown in Figure~\ref{plano}. The
action of $a^{\dagger}_{\ell-1}$ reverses the previous result;
this is represented by $(s,\ell) \rightarrow (s+1,\ell -1)$.
Notice that a particular situation occurs for $\ell =0$: The
R.H.S. relationship (\ref{intertwin1}) prohibits the action of
$a^{\dagger}_0$ on any eigenfunction $\phi^{(0)}_s$ of $H_0$
because it produces the forbidden value $\ell = -1$. However,
function (\ref{sol1e}) can be also derived by applying the
appropriate product of annihilation operators $\{ a_{\ell} \}$ on
a convenient excited state of $H_0$. The straightforward
calculation leads to
\be
\phi^{(\ell)}_s = C^{(\ell)}_s \left( \prod_{k=0}^{\ell -1}
a_{\ell -1 -k} \right) \phi^{(0)}_{s + \ell}, \qquad \ell \neq 0,
\quad s= 0,1,\ldots
\label{ele}
\ee
Another canonical factorization, different from
(\ref{factor1}) and (\ref{factor2b}), is still possible for
$H_{\ell}$:
\be
H_{\ell}= b_{\ell} \, b^{\dagger}_{\ell} + \kappa_{\ell}, \qquad
b^{\dagger}_{\ell}:= -\frac{d}{dr} + \gamma(r,\ell), \qquad
(b^{\dagger}_{\ell})^{\dagger} = b_{\ell}
\label{bes}
\ee
where the factorization constant $\kappa_{\ell}$ is to be fixed
and the superpotential $\gamma(r,\ell)$ fulfills
\be
\gamma' + \gamma^2 = V_{\ell} - \kappa_{\ell}.
\label{rica2}
\ee
The immediate solution of this last equation reads
\be
\gamma = r + \frac{\ell +1}{r}, \qquad \kappa_{\ell} =
-\varepsilon_{\ell} = -(2\ell +3).
\label{supergama}
\ee
Hence, the reverse product (\ref{bes}) is given by
\be
H_{\ell +1} = b^{\dagger}_{\ell} \, b_{\ell} - \theta_{\ell}
\label{factor3}
\ee
and the new intertwining relationships are
\be
b_{\ell -1} \, (H_{\ell} -2) = H_{\ell - 1}\, b_{\ell -1}, \qquad
b^{\dagger}_{\ell} \, (H_{\ell} +2) = H_{\ell +1}\,
b^{\dagger}_{\ell}.
\label{intertwin2}
\ee
The action of $b^{\dagger}_{\ell}$ on the eigenfunction
$\phi^{(\ell)}_s$ of $H_{\ell}$ with eigenvalue $E^{(\ell)}_s$
produces a square-integrable eigen-solution of $H_{\ell +1}$
belonging to the eigenvalue $E^{(\ell)}_s +2 = 2(2s+\ell+1)+3
\equiv E^{(\ell+1)}_s$. Notice that the number of nodes is
preserved; we write $\phi^{(\ell+1)}_s = C^{(\ell+1)}_s\,
b^{\dagger}_{\ell} \, \phi^{(\ell)}_s$. Such an operation is
represented in Figure~\ref{plano} as the horizontal displacement
$(s,\ell) \rightarrow (s,\ell + 1)$. On the other hand, the
function $\phi^{(\ell-1)}_s = C^{(\ell-1)}_s\, b_{\ell -1} \,
\phi^{(\ell)}_s$ is eigenfunction of $H_{\ell -1}$ with eigenvalue
$E^{(\ell-1)}_s = E^{(\ell)}_s -2$. Thus, the action of
$b_{\ell-1}$ is represented by the displacement $(s,\ell)
\rightarrow (s,\ell - 1)$ in the $(s\times\ell)$--plane, as it is
also shown in Figure~\ref{plano}.

Expressions (\ref{factor1}), (\ref{factor2b}), (\ref{bes}) and
({\ref{factor3}) recover the canonical factorizations of
$H_{\ell}$ reported by Fern\'andez, Negro and del~Olmo in 1996
\cite{Fer96} (see also \cite{Sta98,Liu98} and references quoted
therein). As we have shown, the four canonical ladder operators
$a_{\ell}$, $a^{\dagger}_{\ell}$, $b_{\ell}$,
$b^{\dagger}_{\ell}$, induce displacements in the
$(s\times\ell)$--plane by either creating (annihilating) a node in
the wave-function at the cost of decreasing (increasing) the
azimuthal quantum number in one unit or by preserving the number
of nodes but changing the value of $\ell$ in one unit. Although
all these transformations involve a horizontal component in the
displacements of the point $(s,\ell)$, purely vertical
displacements are also possible. Indeed, we have a pair of
mutually adjoint second order differential operators defined as
follows:
\be
\begin{array}{rl}
(S_{\ell}^{\dagger})^{\dagger}= S_{\ell} \!\!\! & := a_{\ell -1}
\, b_{\ell -1} = b_{\ell} \,
a_{\ell}\\[2ex]
& = -H_{\ell} + 2r \frac{d}{dr} + 2r^2 +1.
\end{array}
\label{ene}
\ee
These operators increase or decrease in one unit the number of
nodes in the wave-functions
\be
S^{\dagger}_{\ell}\, \phi^{(\ell)}_s \propto \phi^{(\ell)}_{s+1},
\qquad S_{\ell}\, \phi^{(\ell)}_s \propto \phi^{(\ell)}_{s-1},
\label{ese}
\ee
and intertwine eigenfunctions of $H_{\ell}$ for which the energy
differs in four units
\be
S_{\ell} \, H_{\ell} = (H_{\ell} +4) S_{\ell}, \qquad
S^{\dagger}_{\ell} \, H_{\ell} = (H_{\ell} -4) S^{\dagger}_{\ell}.
\label{eseint}
\ee
Hence, we get the following commutation rules
\be
[S_{\ell}, H_{\ell}] = 4 S_{\ell}, \qquad [S^{\dagger}_{\ell},
H_{\ell}] = - 4 S^{\dagger}_{\ell}.
\label{conmuta1}
\ee
It is clear that the Hamiltonian $H_{\ell}$ is not factorized by
$S_{\ell}$ and $S^{\dagger}_{\ell}$:
\be
S_{\ell}\, S^{\dagger}_{\ell} = (H_{\ell} - \varepsilon_{\ell} +
4)(H_{\ell} +\varepsilon_{\ell}), \qquad S^{\dagger}_{\ell}\,
S_{\ell} = (H_{\ell} + \varepsilon_{\ell} - 4 )(H_{\ell} -
\varepsilon_{\ell}).
\label{ese2}
\ee
However, there is a simple commutation rule
\be
[S_{\ell}, S^{\dagger}_{\ell}] = 8 H_{\ell}.
\label{conmuta2}
\ee
Figure~\ref{crea} shows the schematic representation of two
arbitrary but contiguous energy levels of $H_{\ell}$; the
corresponding points of the $(s \times \ell)$--plane are also
depicted. As indicated, the operator $S^{\dagger}_{\ell}$ scales
up the energy to the closer excited one and $S^{\dagger}_{\ell}$
operates in the opposite direction.

\medskip
%%%%%%%%%%%%%%%%%%%%%%%%%%%%%%
\begin{figure}[ht]
\begin{center}
\includegraphics[height=.20\textheight]{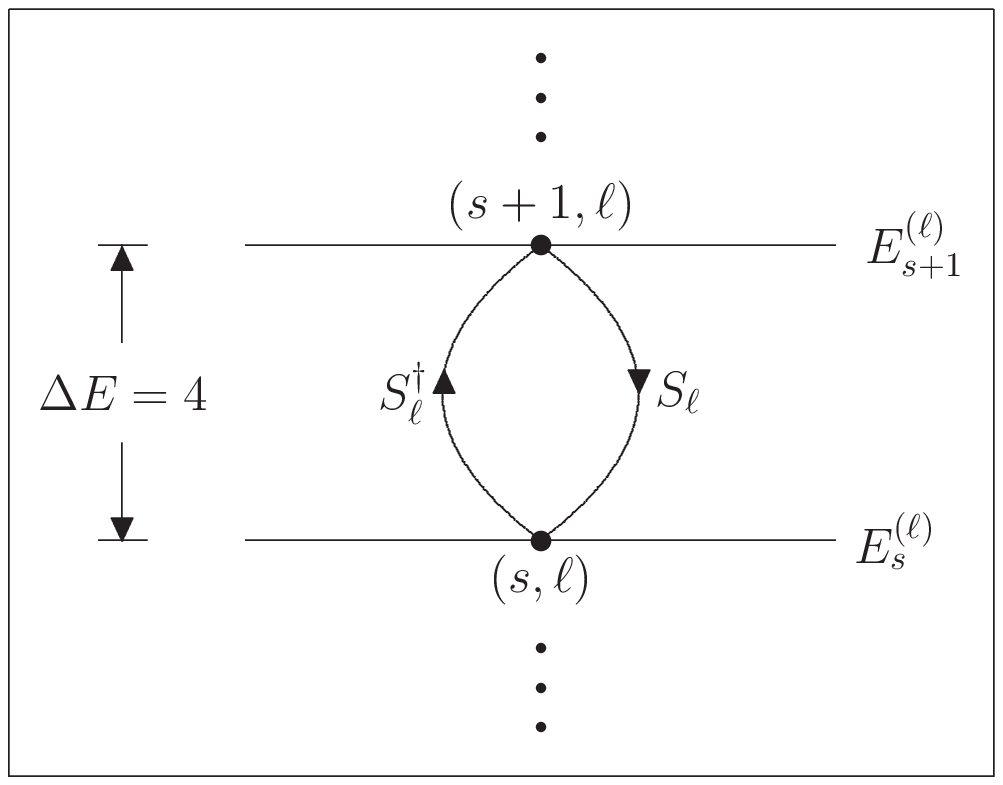}
\caption{\label{crea} The ladder operators $S^{\dagger}_{\ell}$
and $S_{\ell}$ respectively create and annihilate nodes in the
eigenfunctions of $H_{\ell}$. Notice that displacements in the $(s
\times \ell)$--plane are purely vertical, thus the value of the
azimuthal quantum number $\ell$ is fixed.}
\end{center}
\end{figure}
%%%%%%%%%%%%%%%%%%%%%%%%%%%%%%

%%----------------------------------------------------->{Section}
\section{Complex Factorization}

\vskip12pt

In this section we are going to go a step further in the
factorization of Hamiltonian (\ref{ham1}). Let us analyze the
product
\be
H_{\ell} = A_{\ell} \, B_{\ell} + \epsilon_{\ell}
\label{factor4}
\ee
where $\epsilon_{\ell}$ is now a complex factorization constant
with non-trivial imaginary part (from now on $\Re(z)$ and $\Im(z)$
respectively stand for the real and imaginary parts of $z \in
\mathbb{C}$ while $\overline z$ corresponds to its complex
conjugate). The new factorizing operators are established in the
context of a `refined factorization' \cite{Neg00c} to read
\be
A_{\ell} =-\frac{d}{dr} + \beta(r,\ell), \qquad B_{\ell} =
\frac{d}{dr} + \beta(r,\ell)
\label{factor4a}
\ee
with $\beta$ a complex function satisfying the Riccati equation
\be
-\beta' + \beta^2 = V_{\ell} - \epsilon_{\ell}.
\label{ric3}
\ee
The first aspect which distinguishes (\ref{factor4}) from
(\ref{factor1}) is that $A_{\ell}$ and $B_{\ell}$ are not mutually
adjoint in the Hilbert space ${\cal H} = L^2(D_V)$ spanned by the
vectors (\ref{sol1b}). However, as $H_{\ell}$ is selfadjoint,
$H_{\ell} = B^{\dagger}_{\ell} \, A^{\dagger}_{\ell} +
\overline{\epsilon}_{\ell}$ must be true. The straightforward
calculation shows that this last expression leads to the complex
conjugation of equation (\ref{ric3}), the solution of which is
immediate since $V_{\ell}(r)$ is real.

As in the previous section, the factors in (\ref{factor4}) lead to
solutions of the involved eigenvalue problem. In particular, the
function $u(r,\ell)$, annihilated by $B_{\ell}$, is also solution
of the Schr\"odinger equation
\be
H_{\ell} \, u = \epsilon_{\ell} \, u.
\label{uschro}
\ee
Since $\epsilon_{\ell}$ is complex and $H_{\ell}$ is an Hermitian
operator, it is clear that $u$ is not square-integrable in $D_V$.
However, this function is very useful because the non-linear
Riccati equation (\ref{ric3}) can be mapped into (\ref{uschro}) by
means of the logarithmic derivative $\beta = -\frac{d}{dr} \ln u$.
The appropriate transformations allow to express the solutions of
(\ref{uschro}) in terms of confluent hypergeometric functions
${}_1F_1(a,c;z)$ (see e.g. \cite{Neg00a,Neg00b}). In particular,
we shall use
\be
u(r,\ell) = c_{\ell} \, r^{\ell +1} e^{-r^2/2} {}_1F_1 \left(
\frac{\ell}{2} + \frac34 -\frac{\epsilon_{\ell}}{4}, \ell +
\frac32; r^2 \right)
\label{u}
\ee
with $c_{\ell}$ an arbitrary integration constant which will be
fixed as 1. Hence, we have
\be
\beta(r,\ell) = r -\frac{\ell +1}{r} - 2r \,\left( \frac{2\ell + 3 -
\epsilon_{\ell}}{4\ell + 6} \right) \, \frac{{}_1F_1 \left(
\frac{\ell}{2} + \frac74 -\frac{\epsilon_{\ell}}{4}, \ell +
\frac52; r^2 \right)}{{}_1F_1 \left( \frac{\ell}{2} + \frac34
-\frac{\epsilon_{\ell}}{4}, \ell + \frac32; r^2 \right)}.
\label{superb}
\ee
The well known analytical properties of the ${}_1F_1$ functions
lead to
\be
\beta \approx \left\{
\begin{array}{cl}
-\frac{\ell+1}{r}  & r \rightarrow 0\\[1ex]
-r & r \rightarrow +\infty
\end{array}
\right.
\label{superb2}
\ee
The reverse product (\ref{factor4}) produces a new second order
differential operator
\be
B_{\ell} \, A_{\ell} + \epsilon_{\ell} = H_{\ell} + 2 \beta'
\label{factor5}
\ee
which is a non-Hermitian Hamiltonian in $\cal{H}$ since $\beta$ is
complex. The new potential
\be
v(r, \ell, \epsilon_{\ell}):= V_{\ell}(r) + 2 \beta'(r,\ell)
\label{factor5a}
\ee
is mainly a real function at the edges of $D_V$:
\be
\left\{
\begin{array}{l}
v(r \rightarrow 0, \ell, \epsilon_{\ell}) = V_{\ell + 1}(r
\rightarrow 0) +
i\, \frac{4 \Im (\epsilon_{\ell})}{4\ell +6}\\
\\
v(r \rightarrow \infty, \ell, \epsilon_{\ell}) = V_{\ell}(r
\rightarrow \infty)-2
\end{array}
\right.
\label{newpot}
\ee

%%%%%%%%%%%%%%%%%%%%%%%%%%%%%%
\begin{figure}[ht]
\begin{center}
\includegraphics[height=4cm]{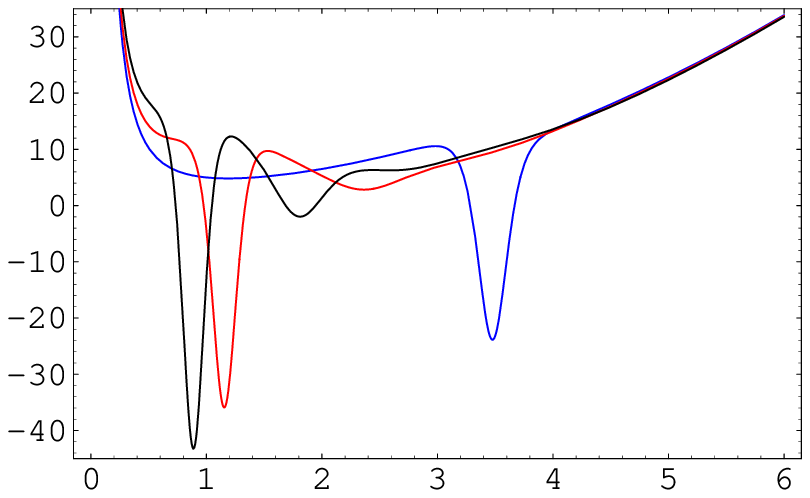} \hskip1cm
\includegraphics[height=4cm]{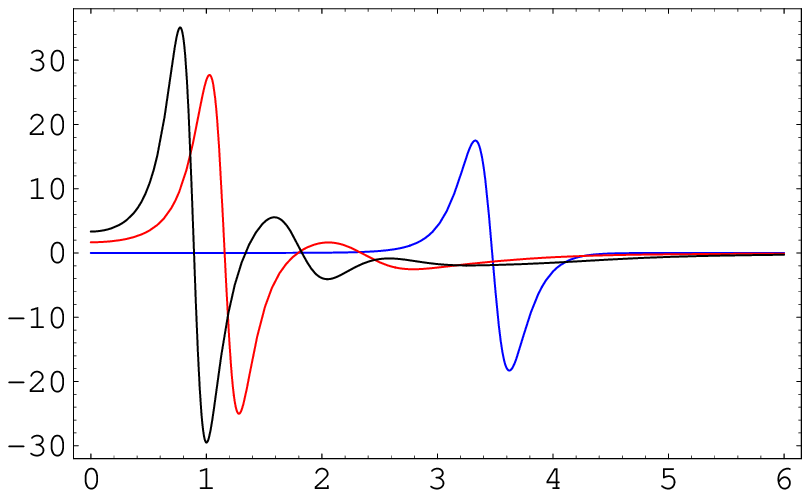}
\caption{\label{fig2} Real (left) and imaginary (right) parts of
the complex potential (\ref{factor5a}) obtained from $V_0(r)$ with
a factorization constant $\epsilon_0= 3 +i\, 10^{-3}$ (blue
curve), $\epsilon_0= 7 +i\, \frac52$ (red curve) and $\epsilon_0=
11 +i5$ (black curve). Observe that greater values of $\Re
(\epsilon_0)$ induce larger displacements to the left in the
deformation}
\end{center}
\end{figure}
%%%%%%%%%%%%%%%%%%%%%%%%%%%%%%

\noindent
Figure~\ref{fig2} shows the global aspect of the new potentials
$v(r,\ell,\epsilon_{\ell})$ for $\ell =0$ and several values of
$\epsilon_0$. Notice that the Darboux-deformation (\ref{factor5a})
of the initial potential $V_{\ell}$ is characterized by a zone $M
\subset D_V$ surrounding the origin. Namely, $M =[0, r_0)$, with
$r_0$ a finite number depending on $\ell$ and $\epsilon_{\ell}$.
In this {\em deformation zone} $M$, the function $\Re (v)$ is
merely a distortion of $V_{\ell +1}$ while $\Im (v)$ is a
nontrivial position-dependent function. For the sake of notation,
we shall write $v_{\ell+1}$ for the complex potential defined in
(\ref{factor5a}). Thereby, the non-Hermitian Hamiltonian
(\ref{factor5}) reads
\be
h_{\ell+1} = B_{\ell} \, A_{\ell} + \epsilon_{\ell} =
-\frac{d^2}{dr^2} + v_{\ell+1}.
\label{factor5b}
\ee
Once we have obtained the factorizations (\ref{factor4}) and
(\ref{factor5b}), it is immediate to arrive at the following
intertwining relationships:
\be
B_{\ell} \, H_{\ell} = h_{\ell+1} \, B_{\ell}, \qquad A_{\ell} \,
h_{\ell+1} = H_{\ell} \, A_{\ell}.
\label{intertwin2a}
\ee
Thus, the action of $B_{\ell}$ on $\phi^{(\ell)}_s$ gives a
function $\psi^{(\ell+1)}_s (r,\epsilon_{\ell}) \propto B_{\ell}
\, \phi^{(\ell)}_s(r)$, which satisfies the Schr\"odinger equation
\be
h_{\ell+1} \, \psi^{(\ell +1)}_s = E^{(\ell)}_s \, \psi^{(\ell
+1)}_s.
\label{hschro1}
\ee

%\medskip
%%%%%%%%%%%%%%%%%%%%%%%%%%%%%%
\begin{figure}[ht]
\begin{center}
\includegraphics[height=4cm]{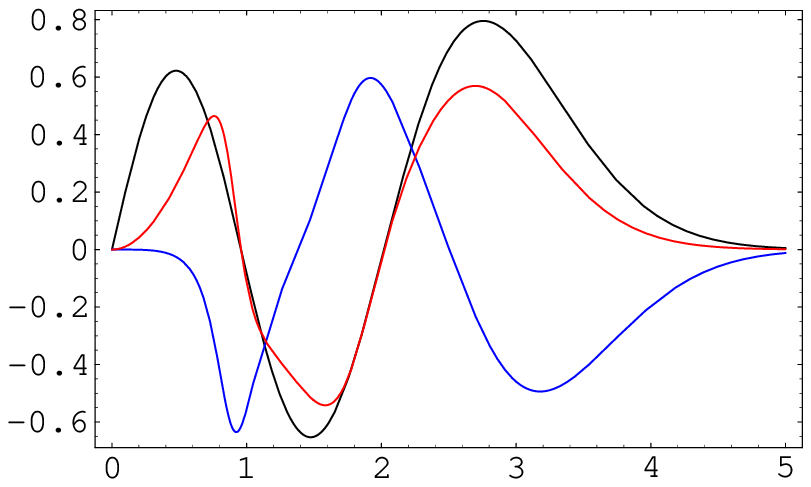} \hskip1cm
\includegraphics[height=4cm]{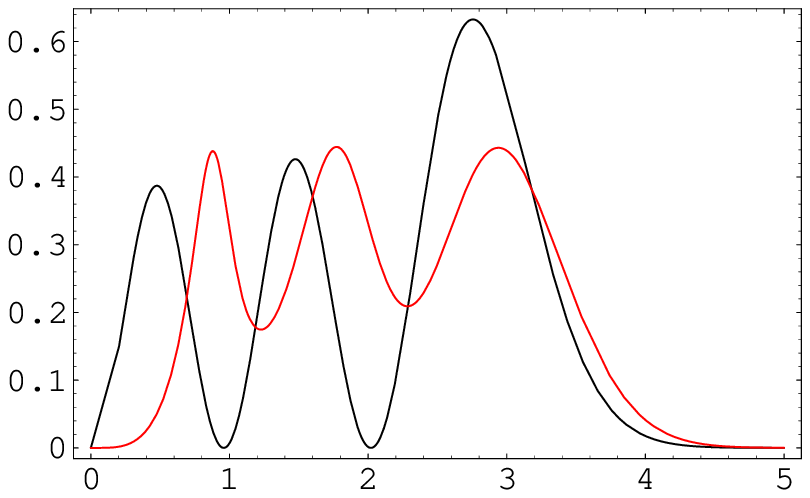}
\caption{\label{fig3} Left: The second excited eigenstate
$\phi^{(0)}_2(r)$ of $H_0$ (black curve) and the real (blue curve)
and imaginary (red curve) parts of the corresponding
Darboux-deformation $\psi^{(1)}_2(r,\epsilon_0)$ for $\epsilon_0 =
11 + i\, 5$. Right: The Born probability densities of
$\phi^{(0)}_2(r)$ (black curve) and $\psi^{(1)}_2(r,\epsilon_0)$.}
\end{center}
\end{figure}
%%%%%%%%%%%%%%%%%%%%%%%%%%%%%%

\noindent
The straightforward calculation shows that this new function
$\psi^{(\ell +1)}_s$ is square-integrable in $D_V$.
Figure~\ref{fig3} depicts the initial eigenfunction
$\phi^{(\ell)}_s$ corresponding to $\ell =0$, $s=2$, as well as
its Darboux-deformation $\psi^{(1)}_2 (r,\epsilon_0 = 11 + i\,
5)$. The function $\phi^{(0)}_2$ corresponds to the second excited
state of $H_0$, accordingly it shows two nodes in $D_V$. However,
although $\psi^{(1)}_2$ corresponds to the second excited energy
$E^{(0)}_2=11$ of $h_1$, it is notable that this function is free
of nodes. Thereby, the usual correspondence between the number of
nodes and the level of excitation of a given energy eigenstate is
missing in the new functions. Despite this fact, the normalized
function $\vert \psi^{(\ell +1)}_s \vert^2$ can be put in
correspondence with the Born's probability density.

In the previous section we realize that the action of the
canonical factorizing operators $a_{\ell}$ and
$a^{\dagger}_{\ell}$ on a given physical state $\phi^{(\ell)}_s$
respectively reduces and increases in one unit the number of nodes
($b_{\ell}$ and $b^{\dagger}_{\ell}$, on the other hand, do not
change the parameter $s$). In contrast, nodes disappear in the
complex Darboux-deformations of physical states, as established by
the set of equations (\ref{factor4a}), (\ref{superb}) and
(\ref{intertwin2a}). In this way, the parameter $s$ in
$\psi^{(\ell +1)}_s$ is nothing but the heritage of the number of
nodes of the initial wave-function $\phi^{(\ell)}_s$. Another
important consequence of the absence of this kind of zeros is that
the new set of square-integrable functions $\{ \psi^{(\ell +1)}_s
\}$ is not orthogonal. However, it is also possible to construct
ladder operators which increase or decrease the energy of this
system.

%\medskip
%%%%%%%%%%%%%%%%%%%%%%%%%%%%%%
\begin{figure}[ht]
\begin{center}
\includegraphics[height=.20\textheight]{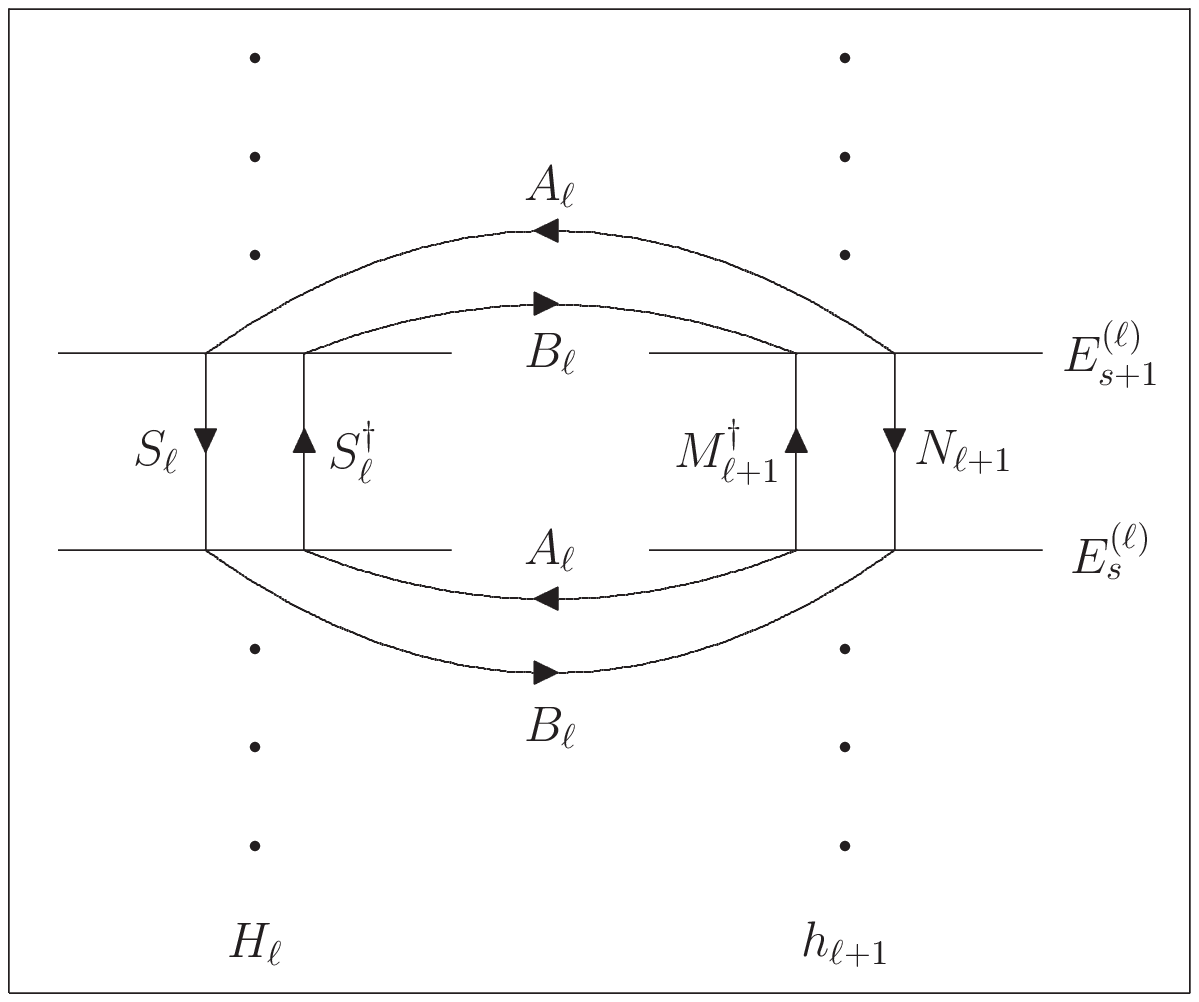}
\caption{\label{ladder} Schematic representation of the
intertwining operators $A_{\ell}$ and $B_{\ell}$, as they are
applied on the energy levels of $H_{\ell}$ and $h_{\ell +1}$. The
construction of the creation and annihilation operators of
$h_{\ell+1}$ is respectively depicted as the composites
$M^{\dagger}_{\ell+1} = B_{\ell} \, S^{\dagger}_{\ell} \,
A_{\ell}$ and $N_{\ell+1} = B_{\ell} \, S_{\ell} \, A_{\ell}$.}
\end{center}
\end{figure}
%%%%%%%%%%%%%%%%%%%%%%%%%%%%%%

\noindent
In analogy with (\ref{eseint}), let us analyze the intertwining
relationships
\be
N_{\ell+1} \, (h_{\ell+1} -4) = h_{\ell+1} N_{\ell+1}, \qquad
M^{\dagger}_{\ell+1} \, (h_{\ell+1}+4) = h_{\ell+1}
M^{\dagger}_{\ell+1}.
\label{mint}
\ee
The above equations give place to the commutation rules
\be
[N_{\ell+1}, h_{\ell+1}] = 4 N_{\ell+1}, \qquad
[M^{\dagger}_{\ell+1}, h_{\ell+1}] = -4 M^{\dagger}_{\ell+1}
\label{conmuta3}
\ee
which are the same as those fulfilled by $H_{\ell}$,
$S^{\dagger}_{\ell}$ and $S_{\ell}$ in (\ref{conmuta1}). Thus,
$M^{\dagger}_{\ell+1}$ and $N_{\ell+1}$ should respectively play
the role of the raising and lowering operators for the eigenvalues
of $h_{\ell +1}$. From (\ref{mint}), it is clear that $N_{\ell+1}
\, \psi^{(\ell +1)}_s$ is eigenfunction of $h_{\ell +1}$ with
eigenvalue $E^{(\ell)}_s-4 = E^{(\ell)}_{s-1}$. Thus, $N_{\ell+1}$
lowers the energy of the system in four units by annihilating one
unit in the parameter $s$. Hence, we write $\psi^{(\ell +1)}_{s-1}
\propto N_{\ell+1} \, \psi^{(\ell +1)}_s$. In a similar way we get
$\psi^{(\ell +1)}_{s+1} \propto M^{\dagger}_{\ell+1} \,
\psi^{(\ell +1)}_s$. Let us remember that $S_{\ell}$ and
$S^{\dagger}_{\ell}$ were constructed in terms of vertical
displacements of the points in the $(s \times \ell)$--plane. In
the present case, we have to solve the algebraic equations
(\ref{mint}--\ref{conmuta3}) and Figure~\ref{ladder} is of special
utility. The simplest solution is given by the following products
\be
M^{\dagger}_{\ell+1} = B_{\ell} \, S^{\dagger}_{\ell} \, A_{\ell},
\qquad N_{\ell+1} = B_{\ell} \, S_{\ell} \, A_{\ell}, \qquad
(M^{\dagger}_{\ell+1})^{\dagger} \neq N_{\ell+1},
\label{products}
\ee
which can be verified by algebraic operations. Since
$M^{\dagger}_{\ell+1}$ and $N_{\ell+1}$ are fourth order
differential operators, they do not factorize the Hamiltonian
$h_{\ell}$. The corresponding products are fourth degree
polynomials factorized as follows
\be
\begin{array}{c}
M^{\dagger}_{\ell+1} \, N_{\ell+1} = (h_{\ell+1} - \epsilon_{\ell}
-4)(h_{\ell+1} + 2\ell -1)(h_{\ell+1} -2\ell -3)(h_{\ell+1} -
\epsilon_{\ell})\\
\\
N_{\ell+1} \, M^{\dagger}_{\ell+1} = (h_{\ell+1} - \epsilon_{\ell}
+4)(h_{\ell+1} - 2\ell +1)(h_{\ell+1} +2\ell +3)(h_{\ell+1} -
\epsilon_{\ell})
\end{array}
\label{poli}
\ee
Finally, it is remarkable that, in contrast with
$S^{\dagger}_{\ell}$ and $S_{\ell}$, the ladder operators
$M^{\dagger}_{\ell+1}$ and $N_{\ell+1}$ are not mutually adjoint
(it is a heritage of the factorizing operators $A_{\ell}$ and
$B_{\ell}$). A similar situation gave place to the {\em Distorted
Heisenberg-Weyl Algebra} reported in \cite{Fer94}. Research in
this direction is in progress.

%%----------------------------------------------------->{Section}
\section{Concluding Remarks}

Conventional factorization is useful to solve the eigenvalue
problem of the spherical symmetric harmonic oscillator. The
corresponding Hilbert space ${\cal H} = L^2(D_V)$ is spanned by
the physical solutions which, in turn, are written in terms of
orthonormal associated Laguerre polynomials. A new kind of
factorizing operators has been introduced to express the radial
oscillator Hamiltonian $H_{\ell}$ as the product of two not
mutually adjoint operators plus a complex constant. The reversed
factorization produces a new Hamiltonian which is not Hermitian
under the inner product of the associated Laguerre polynomials.
However, the corresponding eigenvalue equation has been solved for
the radial oscillator energy spectrum. Although the new solutions
are not orthogonal, they are square-integrable in ${\cal H}$.
Hence, the wave-functions so obtained can be put in correspondence
with the Born probability density.

The complex potentials of the non-Hermitian Hamiltonians reported
here are such that their imaginary part is not a simple constant
but a nontrivial function of the position. There is a well defined
subset of $D_V = [0, \infty)$ in which this `complex deformation'
of the radial oscillator potential is relevant. A similar
situation was reported for complex Coulomb-like \cite{Ros03,Ros07}
and complex linear oscillator-like \cite{Fer03} interactions.
Recent results show that, in a `free particle background', this
kind of potentials behaves as an optical device which both
refracts and absorbs light waves \cite{Fer07b}. Then, it is
interesting to analyze the situation in a different background,
namely in the radial oscillator one. On the other hand, in
Section~3, it has been shown that the wave-functions of these
complex potentials do not have a clear correspondence between the
number of nodes and the level of energy excitation. However, our
current studies show that this correspondence should be
established by means of the real and imaginary parts of the new
functions \cite{Cab07a} (see also \cite{Cab07b}). The improvements
of these last results are going to be reported elsewhere.

%%----------------------------------------------------->{Section}
\section*{Acknowledgments.}
The support of CONACyT project 24233-50766-F is acknowledged.

%%----------------------------------------------------->{end body}

\end{document}